\documentclass[12pt]{iopart}
\usepackage{setstack}
\begin{document}
\title{Dirty black holes: Quasinormal modes}
\author{A J M Medved, Damien Martin, and Matt Visser}
\address{School of Mathematical and Computing Sciences,
Victoria University of Wellington, PO Box 600, Wellington, New Zealand}
\eads{\mailto{joey.medved@mcs.vuw.ac.nz}, \mailto{matt.visser@mcs.vuw.ac.nz}}
\begin{abstract}
 
  In this paper, we investigate the asymptotic nature of the
  quasinormal modes for ``dirty'' black holes --- \emph{generic}
  static and spherically symmetric  spacetimes for which a central
  black hole is surrounded by arbitrary ``matter'' fields.  We
  demonstrate that, to the leading asymptotic order, the [imaginary]
  spacing between modes is precisely equal to the surface gravity,
  independent of the specifics of the black hole system.
  
  Our analytical method is based on locating the complex poles in the
  first Born approximation for the scattering amplitude. We first
  verify that our formalism agrees, asymptotically, with previous studies
  on the Schwarzschild black hole.  The analysis is then generalized
  to more exotic black hole geometries.  We also extend considerations
  to spacetimes with two horizons and briefly discuss the
  degenerate-horizon scenario.

\vskip 0.50cm
\noindent
  Dated:  2 Oct 2003; V2: 9 Oct 2003; V3: 3 Dec 2003; \LaTeX-ed \today
\\
arXiv: gr-qc/0310009
\end{abstract}

\maketitle

\def\d{{\mathrm{d}}}
\def\be{\begin{equation}}
\def\ee{\end{equation}}
\def\Im{\hbox{Im}}
\def\Re{\hbox{Re}}
\def\sech{\hbox{sech}}
\def\max{\hbox{max}}
\def\min{\hbox{min}}
\def\half{{1\over2}}
\def\define{\equiv}
\def\Nordstrom{Nordstr\"om}
\def\paragraph#1{ {\noindent{\bf #1}\quad}}
\def\phase{{\,\hbox{phase}\,}}

\section{Introduction}

It has long been known that small perturbations of a black hole
spacetime will evolve according to a one-dimensional Schrodinger-like
equation~\cite{RW}.  Although no (normalizable) bound-state solutions
are known, there is still a very interesting class of solutions,
commonly referred to as {\it quasinormal
  modes}~\cite{Pr,CD,Leaver,Chandra-book,And2}, that can be used as a
``basis'' for investigating the physics.  The premise behind the
quasinormal modes is to impose ``radiation boundary conditions'' at
the edges of the spacetime (typically, spatial infinity and the black
hole horizon); a constraint which necessitates a {\it complex} value
for the frequency~ (and, hence, the ``quasi'' nomenclature)
\cite{Rev}.  As it turns out, this complex frequency can be labeled by
a discrete quantum number ($n=0,1,2,...\;$) and takes on the following
asymptotic form~\cite{XX3}:
\begin{equation}
k_{qnm}(n) =  i\;  n\; \hbox{(gap)} + \hbox{(offset)} + O[n^{-1/2}] 
\qquad {\rm  as} \qquad  n\rightarrow\infty \;,
\label{00}
\end{equation}
where the ``gap'' and ``offset'' are model-dependent, complex
parameters that are determined by the precise form of the ``potential
barrier'' in the Schrodinger equation. 

For illustrative purposes, let us suppose an  asymptotically
flat spacetime, in which case the gap is purely real. 
A useful physical interpretation is that, for a highly damped black
hole (as described by asymptotically large $n$), the real part of the
offset measures the frequency of emitted radiation, whereas the gap
corresponds to quantized increments in the inverse relaxation
time~\cite{Rev}. Hence, one might expect that both of these quantities
should be fundamental properties of a given black hole.  In this
sense, it is interesting to consider the modes for the specific case
of a scalar or gravitational perturbation outside of a Schwarzschild
black hole (as generated numerically by Nollert~\cite{Nol},
Andersson~\cite{And}, and substantiated analytically by Motl and
Neitzke~\cite{Motl,MN,Neitzke}):
\begin{equation}
k_{qnm}(n) =   i{1\over 4Gm}\left(n+{1\over 2}\right)+{\ln 3\over
 8\pi G m}  + O[n^{-1/2}] 
\quad {\rm  as} \quad  n\rightarrow\infty \;,
\label{000}
\end{equation}
where $m$ is the black hole mass and $G$ is Newton's
constant.\footnote{Here and throughout, all other fundamental
  constants are set to unity and a 3+1-dimensional spacetime is
  presumed.}  For future reference, let us re-express this result
directly in terms of the surface gravity at the horizon
($\kappa=1/[4Gm]$ for a Schwarzschild black hole):
\begin{equation}
k_{qnm}(n) =   i\kappa\left(n+{1\over 2}\right)+{\ln 3\over
 2\pi} \; \kappa  + O[n^{-1/2}] 
\qquad {\rm  as} \qquad  n\rightarrow\infty \;.
\label{0000}
\end{equation}

Note that the gap in the Schwarzschild case is precisely the surface
gravity.  One might be reasonably inclined to wonder if this
occurrence is an artifact of a particularly simple model (having only
one dimensional parameter) or, rather, a resilient feature of more
exotic black hole geometries.  Undoubtedly, this is an important
question but, perhaps strangely, there has been limited progress
towards finding the answer.  Studies on quasinormal modes have, for
the most part, been of a highly model-specific nature (either
numerical~\cite{Nol,And,BK,List1} or
analytical~\cite{Motl,MN,Neitzke,AH,List2}).

In the current paper, we directly address the above question with a
\emph{generic} analytical approach that is elegant in both its
simplicity and its general applicability. (This continues a long-term
project of one of the current authors --- if black hole thermodynamics
is as fundamental as expected, then it should be completely
independent of the detailed specifics of the black hole under
consideration. The thermodynamic features of the black hole system
should be \emph{generic} and unaffected by any ``dirt'' that might
surround the black hole~\cite{DBH,DBH2,DBH3}.)  In our analysis, we
utilize the first Born approximation for the scattering
amplitude~\cite{Text} as a means for approximating the reflection
coefficient.  It is then possible to identify the quasinormal modes by
locating the poles in the scattering amplitude.  (See, for
example,~\cite{Neitzke}.) Note that the Born approximation is normally
thought of as a ``high-energy'' approximation. More specifically, it
becomes increasingly accurate as $|k|\to\infty$ (this is undoubtedly
true for real $k$ and, in all likelihood, complex $k$ as well).
Although this approach may appear ``quick and dirty'', it does
certainly confirm the known asymptotic behavior in the Schwarzschild
case.  Moreover, the basic methodology can readily be extended to
generic black hole spacetimes.  We ultimately show that the gap does
indeed equate with the surface gravity; this being a model-independent
feature of black hole geometries.

Before proceeding, let us point out that there has been, quite
recently, substantial discussion within the gravity community
concerning the real part of the quasinormal frequency ({\it i.e.}, the
real part of the ``offset'')~\cite{Dre,Kun,List3}.  This has followed,
to a large degree, from Hod's observation~\cite{Hod} that the offset
can be used to fix, uniquely, the spacing between the black hole area
eigenvalues. (This observation implies that the area of a black hole
horizon can be quantized to give an evenly spaced spectrum; a notion
that was first advocated by Bekenstein~\cite{Bek}.)  Unfortunately, we
are not able to address this sub-leading (offset) term at the level of
the first Born approximation.  Nonetheless, if quasinormal modes do
indeed play some sort of role in quantum gravity (in spite of the fact
that they arise out of a purely classical process), then it becomes an
important matter to not only calculate but physically motivate the
complete spectral form.  We would argue that the findings of this
paper --- specifically, identifying a universal property of
quasinormal mode spectra --- can be viewed as progress in this
direction.

The rest of the paper is organized in the following manner.  The next
section concentrates on the quasinormal modes of the Schwarzschild
black hole, as this (relatively) simple case nicely illustrates the
formalism and allows for a direct comparison with known results.  In
Section \ref{S:III}, we go on to consider the quasinormal modes for
``generic'' black hole spacetimes (the only stipulations being
staticity and spherical symmetry), including those with more than one
horizon.  At the end, we also discuss the conceptual difficulties of a
degenerate-horizon scenario.  Section \ref{S:IV} provides a brief
summary and discussion.

\section{Schwarzschild black holes} 
\label{S:II}

Although our formal techniques have quite general applicability, the
focus will be, for the time being, on the minimally complicated (but
nonetheless interesting) case of a Schwarzschild black hole.  We are,
simplistically speaking, interested in small perturbations of the
spacetime outside of the black hole horizon.  It is therefore
appropriate to begin with a standard Klein--Gordon equation,
\begin{equation}
 {1\over \sqrt{-g}}\partial_{\mu}
\left[\sqrt{-g}\,g^{\mu\nu}\,\partial_{\nu}\right]\Psi(r,t,\theta,\phi)
 =0\;,
\label{1}
\label{E:KG}
\end{equation}
where $\Psi$ is the massless perturbation field and the metric
describes the background (four-dimensional Schwarzschild) spacetime.
That is,
\begin{equation}
ds^2=-\left(1-{2m\over r}\right)dt^2+\left(1-{2m\over r}\right)^{-1}dr^2
+r^2d\Omega^2\;.
\label{2}
\end{equation}
(Note that, for the remainder, we set $G=1$. If need be, $G$ and other
fundamental constants can be easily re-introduced via dimensional
considerations.)

Employing a separation-of-variables technique and writing~\footnote{With
the assumption that the imaginary part of $k$ is positive --- {\it cf}, 
equation (\ref{000}) ---  we can fix the sign in the temporal exponent
  by requiring the modes to be exponentially decaying in time.}
\begin{equation}
\Psi(r,\theta,\phi,t) = {1\over r} \; \psi(r) \; Y_{\ell m}(\theta,\phi)\;
\exp(+ikt) \;,
\end{equation}
one can convert equation (\ref{1}) into a Schrodinger-like equation of
the form~\cite{RW}
\begin{equation}
{\d^2\over \d r_*^2} \psi - V[r(r_*)] \, \psi = - k^2 \psi\;,
\label{2.5}
\end{equation}
where  $r_*$ is
the so-called {\it tortoise} coordinate defined by~\cite{Wheeler}
\begin{equation}
{\d r_*\over\d r} = {1\over1-{2m/ r}} \;.
\label{3}
\end{equation}
For scalar perturbations, the ``scattering potential''  is found to be
\begin{equation}
V(r) = \left(1-{2m\over r}\right)
\left[{\ell(\ell+1)\over r^2} + {2m\over r^3}\right]\;,
\label{4}
\end{equation}
where $\ell$ is the orbital angular momentum ($l=0,1,2,...$) and let
us emphasize that $V(r)$ is a rational polynomial in $r$ but {\it not}
in $r_*$. There is a natural generalization to higher spin fields;
namely~\cite{Zer},
\begin{equation}
V(r) = \left(1-{2m\over r}\right)
\left[{\ell(\ell+1)\over r^2} + {2m(1-j^2)\over r^3}\right]
\label{4j}
\end{equation}
 and, for the physically most relevant cases,
\begin{equation}
1-j^2 = \left\{ \begin{array}{lcll}
1 &   :&\hbox{scalar} & j=0\\
3/4&  :&\hbox{Dirac} &j=1/2\\
0&    :&\hbox{vector}&j=1\\
-5/4& :&\hbox{Rarita--Schwinger}&j=3/2\\
-3&   :&\hbox{gravity}&j=2 \;.
\end{array}
\right.
\end{equation}

In more generic circumstances, the (generalized) tortoise coordinate
will be a complicated and perhaps unsolvable function of $r$.  In the
present case, however, $r_*=r_*(r)$ can readily be obtained from a
straightforward integration of equation (\ref{3}). This process yields
\begin{equation}
r_*(r) = r + 2m \ln\left[{r-2m\over 2m}\right]\;.
\label{5}
\end{equation}
Although this function is not explicitly invertible, we do know that
the region $r\in(2m,\infty)$ maps into the region $r_* \in (-\infty,
+\infty)$. That is to say, the exterior of the black hole maps into
the entire real line, and we effectively have a one-dimensional
scattering problem.

It is natural, at this point, to impose the physical boundary
conditions of purely outgoing plane waves at spatial infinity
($r_*\rightarrow\infty$) and purely ingoing plane waves at the horizon
($r_*\rightarrow -\infty$). Such boundary conditions can only be
realized for \emph{complex} values of $k$, thus leading to the notion
of \emph{quasinormal-mode} solutions to the wave equation~\cite{Pr}.
(For the physical relevance of these modes, see the previous section;
see also~\cite{Rev}.)

For a generic (one-dimensional) scattering problem, it is known that
the quasinormal modes can be identified with the poles of the
reflection coefficient or, equivalently, the scattering amplitude as a
function of complex asymptotic momenta~\cite{Rev}.  Here, we will
approximate the scattering amplitude by way of the first Born
approximation~\cite{Text}.  Generally speaking, this approximation is
obtained from the Fourier transform of the scattering potential with
respect to the momentum transfer. More specifically, we can regard
\begin{equation}
a(k) \propto \int_{-\infty}^{+\infty} V[r(r_*)] \exp[+2ik r_*] \; \d r_*
\label{6}
\end{equation}
as an approximate form of the scattering amplitude. Take note of the
factor of $+2$ in the exponential. This follows from the fact that the
momentum \emph{transfer} is always minus twice the \emph{incident} momentum
in a one-dimensional scattering:
\begin{equation}
\vec q=\vec k_f-\vec k_i = (-\vec k_i)-\vec k_i = -2 \; \vec k_i \;.
\end{equation}

Our objective is now clear: identify the poles in equation (\ref{6})
with $V(r)$ as given by equation (\ref{4}).  As an initial step, let
us perform a change of variables,
\begin{equation}
a(k) \propto 
\int_{2m}^{+\infty} V[r] \exp[2ik r_*(r)] \; {\d r_*\over\d r} \; \d r \;,
\label{7}
\end{equation}
and then incorporate  equations (\ref{3}) and (\ref{5}) to give
\begin{equation}
a(k) \propto \int_{2m}^{+\infty} 
V[r] \; {r\over 2m} \; \exp[2ik r] \; 
\left[{r-2m\over2m}\right]^{i4mk-1} \; \d r \;.
\label{8}
\end{equation}
A trivial shift in the integration variable then leads to
\begin{equation}
\fl
a(k) \propto \int_{0}^{+\infty} 
V[2m+r] \; {2m+r\over 2m} \;\exp[4ik m]  \exp[2ik r] \; 
\left[{r\over2m}\right]^{i4mk-1} \; \d r \;.
\label{9}
\end{equation}

For future convenience, let us now define
\begin{equation} 
z \equiv -2ikr  \qquad \hbox{that is} \qquad  r = iz/(2k) \; , 
\label{10}
\end{equation}
and so obtain
\begin{eqnarray}
a(k) &\propto& {1\over 2k}\left[{i\over4mk}\right]^{i4mk}
\exp\left[4ikm\right]
\label{11} \\ 
&&\times
 \int_{0}^{+\infty} 
V\left[2m+{iz\over 2k}\right] \; \left(4mk+iz\right) \;  \exp[-z] \;  
z^{i4mk-1} \; \d z.
\nonumber
\end{eqnarray}
At this point, one can, with a brief inspection of the integrand,
anticipate the presence of a linear combination of Gamma
functions.\footnote{Keep in mind the primary definition of the Gamma
  function for positive real $n$: $\Gamma(n)=\int_{0}^{\infty}
  s^{n-1}\;\exp\left[-s\right]\;ds$. Its analytic continuation to the
  complex plane has poles at all non-positive integers.}  For
instance, a (hypothetical) constant term in $V(z)$ would necessitate
contributions that go as $\Gamma\left(i4mk\right)$ and
$\Gamma\left(i4mk+1\right)$.  The identification of poles then becomes
a trivial exercise; in this case, $k_{qnm}(n) = in/(4m)\;$, where $n$
is any non-negative integer.

Let us now be more precise and recall our specific form (\ref{4}) for
the scattering potential. It follows that
\begin{equation}
V\left[2m+{iz\over 2k}\right]
 = {iz\over 4mk+iz}
\left[{\ell(\ell+1)\over \left(2m+{iz\over 2k}\right)^2} + 
{2m(1-j^2)\over \left(2m+{iz\over 2k}\right)^3}\right] \;.
\label{12}
\end{equation}
Substituting into equation (\ref{11}), we then have
\begin{eqnarray}
a(k) &\propto& {i\over 2k}\;  \left[{i\over4mk}\right]^{i4mk} 
 \;\exp[4ikm]
\label{13} 
\\
&&\times
\;  \int_{0}^{+\infty} 
\left[{\ell(\ell+1)\over \left(2m+i{z\over 2k}\right)^2} + 
{2m(1-j^2)\over \left(2m+i{z\over 2k}\right)^3}\right]
\exp[-z] \; 
z^{i4mk} \; \d z \;.
\nonumber
\end{eqnarray}
Since the only immediate concern is the location of the poles, it is
sufficient, for our purposes, to regard $z$ as a small parameter and
Taylor expand the quantity in the square brackets.  Such an expansion
yields a power series in ascending powers of $z$, and it soon becomes
evident that we obtain
\begin{equation}
a(k) \propto \sum_{s=0}^\infty C_s \; \Gamma(i4mk +1 + s) \;,
\label{14}
\end{equation}
where each of the coefficients, $C_s$, is itself a regular and
well-defined quantity \emph{which we do not need to calculate}.

We can now readily locate the poles in the scattering amplitude.  They
occur at
\begin{equation}
i4mk +1 =  - n    \qquad {\rm where} \qquad  n \geq 0 \;,
\label{15}
\end{equation}
so that, up to the validity of the first Born approximation
(which  presumably gets better as $|k|$ becomes larger), we have
\begin{equation}
k_{qnm}(n) = i{n\over 4m} \qquad {\rm where} \qquad  n > 0 \;.
\label{16}
\end{equation}
Let us re-emphasize that the first Born approximation can only be
expected to have validity for very large scattering energies; meaning
that this result can only be trusted when $n>>1$ (or really when
$n\rightarrow\infty$).  Fortunately, this is just the regime we are
interested in.\footnote{Indeed the Born series converges in relatively
  few cases, and is more typically an asymptotic series. Therefore we
  only expect the Born approximation to be sensitive to the leading
  order contributions, at least when applied naively.}

Our outcome for the location of the quasinormal modes agrees,
asymptotically, with the results obtained from both numerical
studies~\cite{Nol,And} and other analytical means~\cite{Motl,MN}.
That is,
\begin{equation}
k_{qnm}(n) = i{n\over 4m} + O[1]   \qquad {\rm when} \qquad n 
\rightarrow\infty.
\label{17}
\end{equation}
This spacing between asymptotic modes, or the {\it gap}, can
alternatively be written as $1/4m=\kappa$, where $\kappa$ is the
surface gravity of the Schwarzschild black hole.  At a first glance,
this could be interpreted as just a happy coincidence; insofar as
there is only one dimensional parameter in the problem.  (That is, the
gap would almost certainly have to be the surface gravity times a
numerical factor.)  It will, however, be shown below that the
asymptotic spacing is \emph{universally} given by the relevant
$\kappa$, irrespective of the details of the black hole spacetime.

Before proceeding, let us point out that this result for the
Schwarzschild gap has been known, even analytically, for quite some
time. For instance, Liu and Mashoon \cite{newLM}, as well as Andersson
\cite{newAn}, have made this observation by identifying the
high-frequency limit of the scattering equation with the confluent
hypergeometric equation. It is, however, unclear how one would
translate this method, or any of the more recent analytical treatments
\cite{Motl,MN}, into a generic setting.

\section{Generic black hole spacetimes} 
\label{S:III}

In this section, the previous formalism will be extended to
``generic'' black hole spacetimes. By generic, we mean static and
spherically symmetric but with an otherwise arbitrary geometry. (So
the black hole can be ``dirty'' in that it may be surrounded by an
arbitrary source of static and spherically symmetric
matter~\cite{DBH,DBH2,DBH3}.)  Eventually, we will elaborate on
spacetimes with two (or possibly more) horizons, with the second
horizon not necessarily being an event horizon {\it per se}
(\emph{e.g.}, the popular model of a Schwarzschild black hole enclosed
by a de~Sitter cosmological horizon, the Kottler or
Schwarzschild--de~Sitter geometry).\footnote{Although the formalism
  can also be extended to multiple-horizon scenarios, it is the
  feeling of the authors that, from an operative viewpoint, a single
  observer in any spherically symmetric geometry would be able to
  deduce the existence of at most two horizons. That is, for any such
  observer, the accessible spacetime will have no more than two
  spatial boundaries and, therefore, no more than two (non-degenerate)
  horizons can ever come into play.  Hence, formal considerations will
  be restricted to spacetimes with one or two horizons.} Horizons are
(for the time being) assumed to be non-degenerate; but the special
case of horizon degeneracy ({\it e.g.}, an extremal
Reissner--\Nordstrom{} black hole) will be discussed at the very end
of the section.

\subsection{Single-Horizon Scenarios}

Given a static spacetime and spherical symmetry, the metric for a
generic black hole can always be expressed, without loss of
generality, in the following manner (see, for
example,~\cite{DBH,wormhole}):
\begin{equation}
\fl
\d s^2 = - e^{-2\phi(r)}\; \left(1-{2m(r)\over r}\right) \d t^2 
+ \left(1-{2m(r)\over r}\right)^{-1} dr^2 + r^2 \d \Omega^2.
\label{18}
\end{equation}
Here $\phi(r)$ is a model-dependent function (related to the
Morris--Thorne ``redshift function''~\cite{Morris-Thorne}), while the
``mass'' parameter $m(r)$ is equivalent to the Morris--Thorne ``shape
function''~\cite{Morris-Thorne}.  Alternatively, we can write
\begin{equation}
\fl
\d s^2 = e^{-2\phi(r)}\; \left(1-{2m(r)\over r}\right) 
\left[ -\d t^2 + {\d r^2\over  e^{-2\phi(r)} 
\left(1-{2m(r)\over r}\right)^2}\right] + r^2 \d \Omega^2 \;,
\label{19}
\end{equation}
which leads, quite naturally, to a generalized tortoise coordinate,
\begin{equation}
{\d r_*\over\d r} = {1\over  e^{-\phi(r)} \left(1-{2m(r)\over  r}\right)}\;;
\label{20}
\end{equation}
and so
\begin{equation}
\d s^2 = e^{-2\phi(r)}\; \left(1-{2m(r)\over r}\right) 
\left[ - \d t^2 + \d r_*^2\right] + r^2 \d \Omega^2 \;.
\label{21}
\end{equation}

Next, let us next revisit the Klein--Gordon equation (\ref{E:KG}) for
a massless perturbation field. In terms of the generic spacetime
described above, this equation expands into
\begin{equation}
\left\{ - \partial_t^2 + {1\over r^2} \partial_{r_*}  r^2 \partial_{r_*} + 
{ e^{-2\phi(r)}\; \left(1-{2m(r)\over r}\right) } 
\; \Delta_2 \right\} \Psi = 0 \;,
\label{22}
\end{equation}
where $\Delta_2$ represents the angular part of the d'Alembertian.  We
can now proceed in the standard way; namely, factoring $\Psi$ into a
temporal part [$\exp({+ikt})$], angular part (the usual spherical
harmonic) and a radial part. One can express the radial part as $r^p\;
u[r(r_*)]$; in which case, the unique choice of $p$ which eliminates
all terms containing $\partial_{r_*} u$ (but not the double
derivatives) happens to be $p=-1$. (This was exactly the same exponent
as occurred in the Schwarzschild case, which is now seen to be
generic.)  Given this choice, a Schrodinger-like form is once again
obtained:
\begin{equation}
{\d^2\over \d r_*^2} u - V[r(r_*)] \, u = - k^2 u,
\label{22.5}
\end{equation}
except that the scattering potential is significantly more complicated
than was found for the Schwarzschild case.  More specifically, some
straightforward calculation yields the following  
result:\footnote{Here, for sake of simplicity, we are considering 
a scalar ($j=0$) perturbation.  It is, however, technically possible
to extend the calculation  to arbitrary $j$, and such
a generalization should  not alter any of our results or conclusions.}
\begin{eqnarray}
\fl
V(r) =  {1\over r}\;(\partial_{r_*}^2  r) + 
e^{-2\phi(r)}\left(1-{2m(r)\over r}\right)
{\ell(\ell+1)\over r^2} 
\\
\lo=
e^{-2\phi(r)}\left(1-{2m(r)\over r}\right)
\left[{\ell(\ell+1)\over r^2}-f(r) 
\right] \;,
\label{23}
\end{eqnarray}
where we have defined
\begin{equation}
f(r)\equiv 
\left(1-{2m(r)\over r}\right) {\phi^{\prime}(r) \over r}
+ {2\over r^2}\left(m^{\prime}(r) -{m(r)\over r}\right) \;,
\label{24}
\end{equation}
and a prime indicates a derivative with respect to $r$.

Provided that $f(r)$ is a well-defined and regular quantity (which
must always be the case since this is equivalent to the tortoise
coordinate being well behaved), it is clear that the generic potential
is qualitatively very similar to that of the Schwarzschild scenario
[\emph{cf}, equation (\ref{4})].  Rather, the sticking point with the
current calculation is that there is, in general, no means of
obtaining a closed form solution for $r_*=r_*(r)$.

We can, however, circumvent the forementioned difficulty by first
taking note of the exact generic expression for the surface
gravity~\cite{DBH},
\begin{equation}
\kappa 
= 
{1\over 2 r_h} e^{-\phi(r_h)} \left[1-8\pi\rho(r_h)\; r_h^2\right] 
=
{1\over 2}\left. 
{\d \over\d r} \left[ e^{-\phi(r)} \; \left(1-{2m(r)\over r} \right) \right] 
\right|_{r_h},
\label{25} 
\end{equation}
where $r=r_h$ indicates the horizon [specified by $2m(r_h)=r_h$] and
$\rho(r)$ is the energy density. Hence we can expand equation
(\ref{20}) to give
\begin{eqnarray}
{\d r_*\over\d r} &=& 
{1\over 2\kappa\left\{(r-r_h)-\alpha(r-r_h)^2 +O[(r-r_h)^3]\right\}} 
\nonumber \\
&=& {1\over 2\kappa (r-r_h)} +{\alpha\over 2\kappa} +O[(r-r_h)] \;,
\label{26}
\end{eqnarray}
where $\alpha$ is a model-dependent constant parameter. (Although such
an expansion is technically valid only in the vicinity of the horizon,
it turns out to be sufficient for the purpose of identifying poles in
the scattering amplitude.  As seen in the previous section, the
location of the poles, at the level of the first Born approximation,
is only sensitive to the near-horizon geometry.)  Moreover, the above
expansion can now be directly integrated to yield
\begin{equation}
r_*={\alpha\over 2\kappa}(r-r_h) 
+{1\over 2\kappa}\ln\left[{r-r_h\over r_h}\right] +O[(r-r_h)^2] \;.
\label{27}
\end{equation}

Beginning with the (approximate) expression for the scattering
amplitude (\ref{6}) and repeating the steps that took us up to
equation (\ref{11}), we now find that (here using the ``convenient''
choice of $z\equiv -i\alpha k r/\kappa$ and neglecting the irrelevant
prefactors)
\begin{equation}
\fl
a(k) \propto 
 \int_{0}^{+\infty} 
V\left[r_h+{iz\kappa\over \alpha k}\right] \; 
\left({k\over\kappa}+iz+O[z^2]\right)
  \exp[-z] \;  \left(1+O[z^2]\right)
z^{i(k/\kappa)-1} \; \d z \;.
\label{28}
\end{equation}
Now what about the scattering potential? An inspection of equation
(\ref{23}) reveals that $V\left[r_h+{iz\kappa/(\alpha k)}\right]$ will
translate into a power series in $z$ times several terms; each of
which contains some (typically negative) power of the argument
$\left[r_h+{iz\kappa/ (\alpha k)}\right]\;$.  As in the previous
section, we can, for our purposes, regard $z$ as small and Taylor
expand appropriately.  Hence, we end up with a complicated power
series in $z$.  Fortunately, the precise details of this expression
are unimportant to us.

Given the previous expression (\ref{28}) and the above discussion, it
becomes clear that
\begin{equation}
a(k) \propto 
 \int_{0}^{+\infty}  
  \exp[-z] \;  
z^{i{k/\kappa}} \; F(z) \; \d z \;,
\label{29}
\end{equation}
where $F(z)$ is some unknown but, in principle, calculable power
series in integer powers of $z$.  [Given that the surface gravity is
non-zero, the leading term in $F$ is of the order $z^0$, as can be
seen by evaluating $\d V(r)/\d r\neq 0$ at the horizon.]

From the above form, it  follows that
\begin{equation}
a(k) \propto \sum_{s=0}^\infty C_s \; 
\Gamma\left(i\;{k\over\kappa} +1+ s\right) \;,
\label{30}
\end{equation}
(As before, when it comes to locating the position of the poles, we do
not need to calculate the coefficients $C_s$.) Hence, the
first-Born-approximated poles are located at
\begin{equation}
k_{qnm}(n) = in\kappa  \qquad {\rm where} \qquad n>0 
\;,
\label{31}
\end{equation}
which is enough to imply that the actual physical poles lie at
\begin{equation}
k_{qnm}(n) = in\kappa + O[1]   \qquad {\rm where} \qquad n>0 
\;,
\label{31b}
\end{equation}
with the constant term becoming (in comparison) irrelevant as
$n\rightarrow\infty$.  This is the main result of the paper and
substantiates our claim at the end of Section 2.

\subsection{Dual-Horizon Scenarios}

Let us now envision an observer ``trapped'' between two horizons; for
instance, the region of spacetime between a Schwarzschild black hole
and a de~Sitter cosmological horizon or the two event horizons in a
Reissner--\Nordstrom{} black hole (although our formalism certainly
allows for much more generic situations).  We propose that it is
straightforward to extend the previous analysis to such situations by
virtue of the following observations:
\begin{enumerate}
\item Given a spherically symmetric and static spacetime, the metric
  can still be cast, without loss of generality, in the form of
  equation (\ref{18}).
\item In the vicinity of any given horizon, the derivative $dr_*/dr$
  (and all related quantities) can be expanded in the form of equation
  (\ref{26}); where the surface gravity ($\kappa$) and horizon
  location ($r_h$) are uniquely defined parameters for the horizon in
  question.
\item At the level of the first Born approximation, the poles in
the scattering amplitude are sensitive only to the near-horizon
geometry (or geometries) of the spacetime.
\end{enumerate}

To further elaborate, starting with the appropriately revised form of
equation (\ref{7}),
\begin{equation}
a(k) \propto 
\int_{r_1}^{r_2} V[r] \exp[2ik r_*(r)] \; {\d r_*\over\d r} \; \d r 
\label{32}
\end{equation}
(where $r_1$ and $r_2$ locate the two horizons in the spacetime), we
can split this integral at some intermediate point (say $r_x$) and
then, by way of the third observation, make the approximation of
extending the integrals to $r_x\rightarrow \pm\infty$.  This procedure
effectively yields two distinct sets of poles, one coming from each
horizon. (These are actually two distinct scattering problems which can be 
distinguished by the orientation of the incident wave.)  
That is, one can anticipate the asymptotic form
(with the surface gravities labeled accordingly)
\begin{equation}
k_{qnm}(n_1) = i\; n_1\; \kappa_1 +O[1] \qquad {\rm or} 
\qquad k_{qnm}(n_2)= i\;n_2\;\kappa_2 +O[1]
\;,
\label{33}
\end{equation}
as $n_{1\;\rm{or}\; 2}\rightarrow\infty$.
Note that the inverse of the surface gravity effectively fixes
the time scale, so that the modes scattered by the
inner/outer horizon ({\it i.e.}, the larger/smaller surface gravity)
will dominate observations  at earlier/later times.
Further note that this
phenomena agrees with analytical estimates, based on the use of the
Poschl--Teller potential, as performed by Suneeta~\cite{Suneeta}.

Implicit in the above discussion is that the two horizons can indeed
be spatially isolated. This immediately rules out spacetimes with
degenerate (or very nearly degenerate) horizons; for instance, an
extremal Reissner--\Nordstrom{} black hole or a Nariai (degenerate
Schwarzschild--de~Sitter) spacetime.  To further complicate matters,
degenerate horizons invariably have a vanishing surface gravity, thus
rendering our previous expansions to be useless in this context.

\subsection{Extremal horizons}

In spite of the difficulties inherent to degenerate-horizon
spacetimes, one possible recourse would be to focus on the
near-horizon form of the second radial derivative of $g_{tt}$ (since
the first derivative vanishes at a degenerate horizon).  That is to
say, we can now expand $dr_*/dr$ as follows:
\begin{eqnarray}
{\d r_*\over\d r} &=& 
\left[
{1\over \alpha (r-r_h)^2+\beta(r-r_h)^3 +\gamma(r-r_h)^4+O[(r-r_h)^5]}
\right] 
\nonumber \\
&=& {1\over \alpha (r-r_h)^2} -{\beta\over \alpha^2(r-r_h)} 
+{\beta^2-\gamma\alpha\over\alpha^3} +O[(r-r_h)] \;,
\label{34}
\end{eqnarray} 
where $\alpha$ (related to the second derivative), $\beta$ and
$\gamma$ are model-dependent constants.  The tortoise coordinate is
then
\begin{equation}
r_* = {-1\over\alpha(r-r_h)}  
-{\beta\over \alpha^2} \; \ln\left[{r-r_h\over r_h}\right] + 
{\beta^2-\gamma\alpha\over\alpha^3}\; (r-r_h) +
O[(r-r_h)^2]
\end{equation}

Now, closely following the previous methodology, we find that [after
shifting the $r$ integration from $(r_h,\infty)$ to $(0,\infty)$] the
scattering amplitude takes on the form
\begin{eqnarray}
a(k) &\propto& \int_{0}^{+\infty} 
V[r_h+r] \; \exp\left[
-{2ik\over\alpha}{1\over r}+ {\beta^2-\gamma\alpha\over\alpha^3}\; 2ikr+O[r^2]
\right] 
\nonumber\\
&&\qquad \times
\;\left({r\over r_h}\right)^{-2ik\beta/\alpha^2} \;
\left({1\over r^2}-{\beta\over\alpha} {1\over r}+O[1]\right)\;  \d r \;.
\label{35}
\end{eqnarray}
With the definition $z\equiv r/r_h$, the above can be rearranged into
\begin{eqnarray}
a(k) &\propto& \int_{0}^{+\infty} 
\exp\left[
-{2ik\over\alpha r_h}{1\over z}
+ {\beta^2-\gamma\alpha\over\alpha^3}\;2ikr_h z
\right]  
\;z^{-2ik\beta/\alpha^2} \; F(z) \; \d z \;,
\label{35b}
\end{eqnarray}
where $F(z)$ is again some (in principle) calculable power series in
$z$ with integer exponents.

An evaluation of the $z$ integral now yields modified Bessel functions
which, apart from the possibility of a trivial pole at $k=0$, do {\it
  not} have poles at finite $|k|$.  We can physically interpret this
finding as follows: When the surface gravity is non-zero, the poles in
complex $k$--space can ultimately be traced back to the small--$z$ (or
near-horizon) behavior of the integrand.  Whereas, in the current case
of a degenerate horizon, the integrand factor $\exp(-ik/r)$ washes out
any possibility of a pole for $k\neq 0$.

Which is to say, by setting $\kappa\to0$ in the general analysis, we
are able to ``predict'' that the quasinormal modes all collapse to
zero momentum.  But what this really means is that the quasinormal
modes (if any exist) do not lie in a region of the complex $k$ plane
where the first Born approximation is trustworthy.  Thus, what we
really expect for extremal black holes is that the quasinormal modes
(if any exist) lie in some bounded region of the complex $k$ plane and
are either finite in number or densely scattered in some bounded
region.

Before concluding, let us point out some discrepancies between our
analysis and other recent works that have considered the quasinormal
modes of a charged black hole; both analytically \cite{Neitzke,AH} and
numerically \cite{BK}.  Firstly, there is evidence that the gap of a
Reissner--\Nordstrom{} black hole --- if it is periodic at all ---
goes, not as the surface gravity at the outer horizon, but rather as a
complicated function of both surface gravities (inner and outer
horizon). Naively, this is not what one would expect from our
findings, given that the interior horizon is not explicitly part of
the scattering problem.  Secondly, the same papers have found that the
highly damped modes of an extremal Reissner--\Nordstrom{} black hole
are formally identical to a Schwarzschild black hole of the same mass.
This is clearly contrary to our arguments above.  It is quite possible
that our methodology breaks down, in some subtle way, in spacetimes
with a ``hidden horizon''; that is, it is feasible that the first Born
approximation is unable to properly account for these added
complexities.  It should be noted, however, that some of these authors
have also commented on interpretative difficulties \cite{AH} and
numerical instabilities \cite{BK} on account of some ``peculiar
features'' \cite{AO} of the quasinormal spectra of charged black
holes.  Moreover, the Schwarzschild limit of their Reissner-Nordstrom
spectrum does not appear to give back the desired result.  We would
suggest that the quasinormal mode problem for this charged model
requires further attention.

\section{Conclusion}
\label{S:IV}

To summarize, we have used a simple analytical method --- based on the
first Born approximation for the scattering amplitude --- to locate
the quasinormal modes for a static, spherically symmetric but
otherwise generic black hole spacetime.  For the very special case of
a Schwarzschild black hole, our methodology was found to agree,
asymptotically, with the results obtained from various numerical and
analytical studies.  Moreover, that the asymptotic spacing or ``gap''
is equivalent to the surface gravity (as is accepted in the
Schwarzschild case) was shown to be a model-independent feature of a
wide range of black hole spacetimes.  We also generalized this outcome
to spacetimes with two horizons and reported some limited progress in
the context of degenerate-horizon geometries.

The virtues of our formal treatment include generality and a
straightforward analytical approach that does not obscure the physical
process being investigated.  Nevertheless, the simplicity of our
method has a price: we have (so far) only been able to locate the
leading term in the asymptotic expansion of the quasinormal mode.
Meanwhile, the next-order term --- that is, the ``offset'' --- has
sparked considerable recent interest because of a conjectural
relationship with the black hole area spectrum.  It is unfortunate
that the Born approximation seems to diverge at higher orders, so that
our approach would have to be significantly modified to reproduce the
entire spectrum.

\section*{Acknowledgments} 

Research supported by the Marsden Fund administered by the New Zealand
Royal Society and by the University Research Fund of Victoria
University.

\section*{References}


\begin{thebibliography}{66}


\bibitem{RW} 
T. Regge and J.A. Wheeler, 
``Stability of a Schwarzschild singularity'',
Phys. Rev. {\bf 108}, 1063 (1957).


\bibitem{Pr} 
W.H. Press, 
``Long wave trains of gravitational waves from a vibrating black hole'', 
Astrophys. J. {\bf 170}, L105 (1971).

\bibitem{CD}
S. Chandrasekhar and S. Detweiler,
``The quasi-normal modes of the Schwarzschild black hole'',
 Proc.\ Roy.\ Lond.\ A {\bf 344}, 444 (1975); 
``On the reflection and transmission of neutrino waves by a Kerr black hole'',
Proc.\ Roy.\ Soc.\ Lond.\ A {\bf 352}, 325 (1977). 

\bibitem{Leaver}
E.W.~Leaver,
``An analytic representation for the quasi normal modes of Kerr black holes'',
Proc.\ Roy.\ Soc.\ Lond.\ A {\bf 402}, 285 (1985).

\bibitem{Chandra-book}
S.~Chandrasekhar,
{\it The mathematical theory of black holes},
(Oxford Science Publications, 1983). 

\bibitem{And2} 
N.~Andersson and B.P.~Jensen, 
``Scattering by black holes'',
 arXiv:gr-qc/0011025. 




\bibitem{Rev} See, for comprehensive reviews: \\
H.-P. Nollert,
``Quasinormal modes: the characteristic sound of black holes 
and neutron stars'',
 Class. Quant. Grav. {\bf 16}, R159 (1999);  
\\
K.D. Kokkotas and B.G. Schmidt, 
``Quasi-Normal modes of black holes and stars'',
Living Rev. Rel. {\bf 2}, 2 (1999) [arXiv:gr-qc/9909058].



\bibitem{XX3} 
A. Bachelot and A. Motet-Bachelot, 
``Resonances of Schwarzschild black holes'', in 
{\it Proceedings of the IV International Conference of Hyperbolic Problems},
ed. Vieweg, (Taosmina, 1992).

\bibitem{Nol} 
H.-P. Nollert,
``Quasinormal modes of Schwarzschild black holes: the determination
of quasinormal frequencies with very large imaginary parts'',
 Phys. Rev. D {\bf 47}, 5253 (1993).


\bibitem{And} 
N. Andersson, ``On the asymptotic distribution of
quasinormal-mode frequencies for Schwarzschild black holes'',
Class. Quant. Grav. {\bf 10}, L61 (1993).

\bibitem{Motl} 
L.~Motl,
``An analytical computation of asymptotic 
Schwarzschild quasinormal  frequencies'',
Adv.\ Theor.\ Math.\ Phys.\  {\bf 6}, 1135 (2003) 
[arXiv:gr-qc/0212096].

\bibitem{MN}
L.~Motl and A.~Neitzke,
``Asymptotic black hole quasinormal frequencies'',
Adv.\ Theor.\ Math.\ Phys.\  {\bf 7}, 307 (2003) 
[arXiv:hep-th/0301173].

\bibitem{Neitzke}
A. Neitzke, ``Greybody factors at large imaginary frequencies'',
arXiv:hep-th/0304080 (2003).


\bibitem{BK}
E. Berti and K.D. Kokkotas, 
``Asymptotic quasinormal modes of Reissner-Nordstrom and Kerr black
holes'', Phys. Rev. D {\bf 68}, 044027 (2003) [arXiv:hep-th/0303029].


\bibitem{List1} 
See, for other recent examples:
\\
K. Glampedakis and N. Andersson, ``Quick and dirty methods for
studying black-hole resonances'', Class. Quant. Grav. {\bf 20},
3441 (2003) [arXiv:gr-qc/0304030];
\\
L.-H. Xue, Z.-X. Shen, B. Wang and R.-K. Su,
``Numerical simulation of quasi-normal modes in time-dependent
background'', arXiv:gr-qc/0304109 (2003);
\\
V. Cardoso, R. Konoplya and J.P.S. Lemos, 
``Quasinormal frequencies of Schwarzschild black holes
in anti-de Sitter spacetimes: a complete study on asymptotic behavior'',
Phys. Rev. D {\bf 68}, 044024 (2003) [arXiv:gr-qc/0305037];
\\
E. Berti, V. Cardoso, K. Kokkotas and H. Onozawa,
``Highly damped quasinormal modes of Kerr black holes'',
arXiv:hep-th/0307013 (2003); 
\\
V. Cardoso, J.P.S Lemos and S. Yoshida, 
``Quasinormal modes of Schwarzschild black holes in four
and higher dimensions'', arXiv:gr-qc/0309112 (2003).


\bibitem{AH}
N. Andersson and C.J. Howls, 
``The asymptotic quasinormal mode spectrum of non-rotating black holes'',
arXiv:gr-qc/0307020 (2003).





\bibitem{List2} 
See, for other recent examples:
\\
V. Cardoso, O.J.C. Dias and J.P.S. Lemos,
``Gravitational Radiation in D-dimensional spacetimes'',
Phys. Rev. D {\bf 67}, 064026 (2003) [arXiv:hep-th/0212168];
\\
V. Cardoso and J.P.S. Lemos, ``Quasinormal modes of the
near extremal Schwarzschild-de Sitter black hole'',
Phys. Rev. D {\bf 67}, 084020 (2003) [arXiv:gr-qc/0301078];
\\
R.A. Konoplya, ``Quasinormal behavior of the D-dimensional Schwarzschild
black hole and higher order WKB approach'',
Phys. Rev. D {\bf 68}, 024018 (2003) [arXiv:gr-qc/0303052];
``Gravitational quasinormal radiation of higher-dimensional
black holes'', arXiv:hep-th/0309030 (2003);
\\
H.-T. Cho, ``Dirac quasi-normal modes in Schwarzschild black
hole spacetimes'', Phys. Rev. D {\bf 68}, 024003 (2003) [arXiv:gr-qc/0303078];
\\ 
A. Maassen van den Brink, ``The WKB analysis of the Regge-Wheeler
equation down in the frequency plane'', arXiv:gr-qc/0303095 (2003);
``Approach to the extremal limit of the Schwarzschild-de Sitter
black hole'', Phys. Rev. D {\bf 68}, 047501 (2003) [arXiv:gr-qc/0304092];
\\
C. Molina, ``Quasinormal modes of d-dimensional spherical black
holes with a near extreme cosmological constant'', 
Phys. Rev. D {\bf 68}, 064007 (2003) [arXiv:gr-qc/0304053];
\\
D. Birmingham, ``Asymptotic quasinormal frequencies of
d-dimensional Schwarzschild black holes'', 
Phys. Lett. B {\bf 569}, 199 (2003) [arXiv:hep-th/0306004];
\\
A. Zhidenko, ``Quasi-normal modes of Schwarzschild-de Sitter black holes'',
arXiv:gr-qc/0307012 (2003);
\\
S. Musiri and G. Siopsis, 
``Perturbative calculation of quasi-normal modes of Schwarzschild
black holes'', arXiv:hep-th/0308168 (2003);
``On quasi-normal modes of Kerr black holes'', arXiv:hep-th/0309227 (2003);
\\
K.H.C. Castello-Branco and E. Abdalla, 
``Analytic determination of the asymptotic quasi-normal mode spectrum
of Schwarzschild-de Sitter black holes'', arXiv:gr-qc/0309090 (2003);
\\
E. Berti, M. Cavaglia and L. Gualtieri,
``Gravitational energy loss in high energy particle collisions:
ultrarelativistic plunge into a multidimensional black hole'',
arXiv:hep-th/0309203 (2003).


\bibitem{DBH} 
M.~Visser,
``Dirty black holes: Thermodynamics and horizon structure'',
Phys.\ Rev.\ D {\bf 46}, 2445 (1992) 
[arXiv:hep-th/9203057].

\bibitem{DBH2} 
M.~Visser,
``Dirty black holes: Entropy versus area'',
Phys.\ Rev.\ D {\bf 48}, 583 (1993)
[arXiv:hep-th/9303029].

\bibitem{DBH3} 
M.~Visser,
``Dirty black holes: Entropy as a surface term'',
Phys.\ Rev.\ D {\bf 48}, 5697 (1993) 
[arXiv:hep-th/9307194].


\bibitem{Text} 
See, for example: \\ 
R. Shankar, {\it Principles of quantum mechanics}, 
(Plenum, New York, 1980).

\bibitem{Dre} 
O.~Dreyer,
``Quasinormal modes, the area spectrum, and black hole entropy'',
Phys.\ Rev.\ Lett.\  {\bf 90}, 081301 (2003)
[arXiv:gr-qc/0211076].


\bibitem{Kun} 
G.~Kunstatter,
``$d$-dimensional black hole entropy spectrum from quasi-normal modes'',
Phys.\ Rev.\ Lett.\  {\bf 90}, 161301 (2003)
[arXiv:gr-qc/0212014].

\bibitem{List3} 
Also see, for instance:
\\
A. Corichi, 
``On quasinormal modes, black hole entropy, and quantum geometry'',
Phys. Rev. D {\bf 67}, 087502 (2003) [arXiv:gr-qc/0212126];
\\
S. Hod, ``Kerr black hole quasinormal frequencies'',
Phys. Rev. D {\bf 67}, 081501 (2003) [arXiv:gr-qc/0301122];
``Asymptotic quasinormal mode spectrum of rotating black holes'',
arXiv:gr-qc/0307060 (2003); 
\\
R.K. Kaul and S.K. Rama, ``Black hole entropy from spin one punctures'',
Phys. Rev. D {\bf 68}, 024001 (2003) [arXiv:gr-qc/0301128];
\\
E. Abdalla, K.H.C. Castello-Branco and A. Lima-Santos,
``Area quantization in quasi-extreme black holes'',
Mod. Phys. Lett. A {\bf 18}, 1435 (2003) [arXiv:gr-qc/0301130];
\\
A.P. Polychronakos, 
``Area spectrum and quasinormal modes of black holes'',
arXiv:hep-th/0304135 (2003);
\\
D. Birmingham, S. Carlip and Y. Chen, 
``Quasinormal modes and black hole quantum mechanics in 2+1 dimensions'',
Class. Quant. Grav. {\bf 20}, L239 (2003) [arXiv:hep-th/0305113];
\\
J. Swain, 
``The Pauli exclusion principle and SU(2) vs. SO(3) in loop quantum gravity'', 
arXiv:gr-qc/0303073 (2003);
\\
J. Oppenheim, 
``The spectrum of quantum black holes and quasinormal modes'',
arXiv:gr-qc/0307089 (2003).
\\
Y.~Ling and H.~B.~Zhang,
``Quasinormal modes prefer supersymmetry?'',
arXiv:gr-qc/0309018.

\bibitem{Hod} 
S.~Hod,
``Bohr's correspondence principle and the 
area spectrum of quantum black  holes'',
Phys.\ Rev.\ Lett.\  {\bf 81},  4293 (1998)
[arXiv:gr-qc/9812002].


\bibitem{Bek} 
J.D.~Bekenstein,
``The Quantum Mass Spectrum Of The Kerr Black Hole'',
Lett.\ Nuovo Cim.\  {\bf 11}, 467 (1974).


\bibitem{Wheeler} 
J.A.~Wheeler,
``Geons'',
Phys. Rev. {\bf 97},  511 (1955).


\bibitem{Zer} F.J. Zerilli, ``Gravitational field of a particle
falling in a Schwarzschild geometry analyzed in tensor harmonics'',
Phys. Rev. D {\bf 2}, 2141 (1970).

\bibitem{newLM} H. Liu and B. Mashoon, ``On the spectrum
of oscillations of  Schwarzschild black holes'',
  Class. Quant. Grav. {\bf 13},
233 (1996).

\bibitem{newAn} N. Andersson, ``Evolving test-fields in a black hole
geometry'',  Phys. Rev. D {\bf 55}, 468 (1997) [arXiv:gr-qc/9607064]. 


\bibitem{wormhole} 
M.~Visser,
{\it Lorentzian Wormholes: From Einstein To Hawking},
(AIP Press, Woodbury, 1995).


\bibitem{Morris-Thorne}
M.S.~Morris and K.S.~Thorne,
``Wormholes in space-time and their use for interstellar travel: 
A tool for teaching general relativity'',
Am.\ J.\ Phys.\  {\bf 56}, 395 (1988).

\bibitem{Suneeta}
V. Suneeta, 
``Quasinormal modes for the SdS black hole: 
an analytical approximation scheme'', 
Phys. Rev. D {\bf 68}, 024020 (2003)
[arXiv:gr-qc/0303114].


\bibitem{AO} N. Andersson and H. Onozawa, 
``Quasinormal modes of nearly extreme Reissner-Nordstrom black holes'',
Phys. Rev. D {\bf 54}, 7470 (1996) [arXiv:gr-qc/9607054].


\end{thebibliography}
\end{document}